\input{aipcheck}

\documentclass[
    ,final            % use final for the camera ready runs
  ]
  {aipproc}

\layoutstyle{6x9}
\usepackage{amsmath}
\begin{document}

\title{NUCLEAR PHOTONICS}


\author{D. Habs}{ 
  address={Ludwig-Maximilians-Universit\"at M\"unchen, D-85748 Garching, Germany\\
 Max Planck Institut f\"ur Quantenoptik, D-85748 Garching, Germany}
}

\author{M.M. G\"unther}{ 
  address={Max Planck Institut f\"ur Quantenoptik, D-85748 Garching, Germany}
}

\author{M. Jentschel}{ 
  address={Institut Laue-Langevin, F-38042 Grenoble}
}

\author{P.G. Thirolf}{
  address={Ludwig-Maximilians-Universit\"at M\"unchen, D-85748 Garching, Germany}
}

  %,altaddress={<author1 address>} % additional visiting address

\begin{abstract}
With the planned new $\gamma$-beam facilities like MEGa-ray at LLNL (USA)
or ELI-NP at Bucharest (Romania) with $10^{13}\gamma$/s and a band width of
$\Delta E_{\gamma}/E_{\gamma}\approx 10^{-3}$, a new era of $\gamma$ beams 
with energies up to 20 MeV comes into operation, compared to the present
world-leading HI$\gamma$S facility at Duke University (USA) with 
$10^{8}\gamma/s$ and $\Delta E_{\gamma}/E_{\gamma}\approx 3 \cdot 10^{-2}$.
In the long run even a seeded quantum FEL for $\gamma$ beams may become
possible, with much higher brilliance and spectral flux.  
At the same time new exciting possibilities open up for focused $\gamma$ beams.
Here we describe a new experiment at the $\gamma$ beam of the ILL reactor 
(Grenoble, France), where we observed for the first time that the index of 
refraction for $\gamma$ beams is determined by virtual pair creation.
Using a combination of refractive and reflective optics, efficient 
monochromators for $\gamma$ beams are being developed. 
Thus, we have to optimize the total system: the $\gamma$-beam facility, 
the $\gamma$-beam optics and $\gamma$ detectors. 
We can trade $\gamma$ intensity for band width, going down to  
$\Delta E_{\gamma}/E_{\gamma}\approx 10^{-6}$ and address 
individual nuclear levels. The term ``nuclear photonics'' stresses the 
importance of nuclear applications. We can address with $\gamma$-beams 
individual nuclear isotopes and not just elements like 
with X-ray beams. Compared to X rays, $\gamma$ beams can penetrate much deeper
into big samples like radioactive waste barrels, motors or 
batteries. We can perform tomography 
and microscopy studies by focusing down to $\mu$m resolution
using Nuclear Resonance Fluorescence (NRF) for detection with eV resolution
and high spatial resolution at the same time.
We discuss the dominating M1 and E1 excitations like the scissors mode, 
two-phonon quadrupole octupole excitations, pygmy dipole excitations 
or giant dipole excitations under the new facet of applications. 
We find many new applications in biomedicine, green energy, 
radioactive waste management or homeland security. Also more brilliant 
secondary beams of neutrons and positrons can be produced.

\end{abstract}

\maketitle

%%%%%%%%%%%%%%%%%%%%%%%%%%%%%%%%%%%%%%%%%%%%
%% MAINMATTER
%%%%%%%%%%%%%%%%%%%%%%%%%%%%%%%%%%%%%%%%%%%%

\section{INTRODUCTION}

Here we consider tunable $\gamma$ beams with typical $\gamma$ energies
from 0.1 - 20 MeV from hard X-rays to $\gamma$ beams. 
Some of the properties of these beams,
like their wave length, or the real part $\delta$ or imaginary part 
$\beta$ of the index of refraction 
$n(E_{\gamma})=1+\delta(E_{\gamma})+ i\cdot \beta(E_{\gamma})$ 
for a typical material like silicon and gold are compiled in Table 1.
The refractive part $\delta=-\frac{\lambda^2 \rho b}{2\pi}$,
where $b$ is the scattering length and $\rho$ the atomic density.
The absorptive part 
$\beta=\frac{\lambda \rho \sigma_a}{2\pi}=\frac{\mu\lambda}{4\pi}$,
where $\sigma_a$ is the absorption cross section and
$\mu$ the absorption coefficient.  
For photons we have the relation between the wavelength $\lambda$
in fm and the photon energy $E$ in MeV
  \begin{equation}
  \label{eq:photon}
\lambda \text{[fm]} =\frac{h c}{E}=\frac{1240}{E[MeV]}
\end{equation}
The new $\gamma$ optics \cite{habs11} with small radii of curvature 
works best for rather brilliant beams with a rather small beam diameter. 
Thus, we have to compare the properties of planned and existing  
beam facilities to the requirements of the $\gamma$ optics. 
 \begin{table}[h!]
\caption{Measured and predicted properties of $\gamma$ rays in silicon and gold.}
\smallskip
\begin{tabular}{|c|c|c|c|c|c|} \hline
 energy & wavelength &$\delta$(Si)    &$\beta$(Si)&$\delta$(Au)&$\beta$(Au) \\ \hline
10 MeV  & 124 fm &$+2\cdot 10^{-10}$   &$6\cdot 10^{-14}$ & $3\cdot (10^-6-10^{-7})$& $5\cdot 10^{-12}$ \\ 
 1 MeV  & 1240 fm&$+1\cdot 10^{-9}$    & $2\cdot 10^{-12}$& $10^{-5}-10^{-6}$       & $2\cdot 10^{-10}$ \\ 
0.1 MeV & 0.124{\AA}&$-2\cdot 10^{-7}$ & $6\cdot 10^{-11}$& $10^{-5}-10^{-6}$       &$1\cdot 10^{-7}$ \\ 
12.4 keV& 1.0{\AA}&$-1\cdot 10^{-6}$   & $6\cdot 10^{-8}$  & $2\cdot 10^{-5}$        &$4\cdot 10^{-6}$\\ \hline
\end{tabular}
\end{table}

In Table 2 we have compiled these properties. 
For the newly planned Compton back-scattering $\gamma$-beam facilities like 
MEGa-ray \cite{barty11} and ELI-NP \cite{ELI-NP11}, we expect to get much more
brilliant and much more intense $\gamma$ beams, compared to the present
worldwide best values of the HI$\gamma$S facility \cite{weller09}.
However, the large distance of about 30~m for HI$\gamma$S from the 
production point of the $\gamma$ beam to the first access 
of $\gamma$ optics results in a large $\gamma$-beam diameter of  1 cm, 
too large for an effective use of $\gamma$ lenses. 
On the other hand, for MEGa-ray and ELI-NP first beam diameters 
of about 50 $\mu$m are accessible for an energy spread of $10^{-3}$, well suited
for $\gamma$ optics. For hard X-ray beams of the ESRF facility (Grenoble) 
at the high energy beam line ID15 typical beam diameters are
less than 1000 $\mu$m, again suited for X-ray optics \cite{snigirev11}. 
The smaller beam size allows for smaller optical elements with less
absorption. Then more elements N can be used behind each other
increasing the deflection. On the other hand, we have rather small opening
angles of the $\gamma$ beams. Thus, a small angle change makes the divergent 
$\gamma$ beam convergent. We have to use rather long flight paths
(several 10 meters) to reach even smaller $\gamma$-beam diameters, which then
make $\gamma$ optics even more effective. Thus, similar to hard X-ray
experiments at large synchrotrons, rather long beam lines are required
for $\gamma$-beam experiments. 

We have to look at the three components: (i) the $\gamma$ beam 
(ii) the $\gamma$ optics (iii) the $\gamma$ detectors as an integral system
and have to tailor them for each experiment. Frequently, there is 
a clear need to reduce the average $\gamma$-beam intensity or the
instantaneous intensity on target, but we should not just reduce it by 
collimators, but we should improve at the same time
the band width from an expected value of $10^{-3}$ to, e.g., $10^{-6}$.
It may be advantageous to have the possibility to gradually zoom
in the $\gamma$ beam resolution, to identify first interesting
spectral structures with lower resolution and then 
to increase the beam resolution accordingly. 

\begin{table} [b!]
\caption{Properties of $\gamma$ beam facilities.}
\smallskip
%\begin{center}
\begin{tabular}{|c|c|c|c|c|c|} \hline
property             & HI$\gamma$S   &   MeGa-ray& ELI-NP & $\gamma$QFEL  & ESRF(ID15)  \\  \hline
$\gamma$ energy [MeV] & 1-160         &0.5-2.7    &0.5-19.6& 0.5-10.0 & 0.1-0.70       \\ 
peak brilliance      &               &           &        &          &                 \\
1/(mm$^2$mrad$^2$s~$0.1\%$ BW)&$10^{15}$&$10^{22}$&$10^{22}$&$10^{32}$&$10^{11}$\\ 
intensity [1/s]       & $10^8$        & $10^{13}$ &$10^{13}$&$10^{13}$ &$10^{9}$\\ 
spect. flux [1/(eV~s)]&$6\cdot 10^{2}$&$1\cdot 10^{6}$&$1\cdot 10^{6}$&$10^{11}$&$ 10^{8}$\\ 
band width $\Delta E_{\gamma}/E_{\gamma}$& $3\%$         & $10^{-3}$ &$10^{-3}$ &$10^{-6}$ &$10^{-3}$ \\ 
opening angle at 1 MeV & $1\cdot 10^{-4}$&$1\cdot 10^{-5}$ &$1\cdot 10^{-5}$&$1\cdot 10^{-6}$ &$5\cdot 10^{-6}$\\
closest beam radius [$\mu$m]&$10^4$  & $10^2$           &$10^2$          &$10$           &$10^3$ \\
repetition rate [1/s]& $3\cdot 10^8$ & $1.2\cdot 10^4$  &$1.2\cdot 10^4$ &$1.2\cdot 10^4$  & $3\cdot 10^8$\\ \hline 
\end{tabular}
%\end{center}
\end{table}

\section{ INTEGRATED $\gamma$ BEAM FACILITIES}

\subsection{$\gamma$ Beams}

For the two planned Compton back-scattering $\gamma$ facilities ELI-NP and 
MEGa-ray we expect $10^{13}\gamma/ s$ for all $\gamma$ energies. 
For Compton back-scattering there is a relation between
$\gamma$ energy $E_{\gamma}$ and the angle $\Theta$ of the outgoing 
photon with respect to the electron beam direction:

\begin{equation}
  \label{eq:compton}
     E_{\gamma}=\frac{4\gamma_e^2 \cdot E_L}{1+
(\gamma_e\Theta)^2+\frac{4\gamma_e E_L}{m_ec^2}}
\end{equation}

$E_L$ is energy of the primary laser photons, $\gamma_e=\frac{E_e}{m_ec^2}$
is the $\gamma$ factor of the electron beam with energy $E_e$ and the
electron rest mass $m_e$. 
Thus, for $\Theta=1/\gamma_e$ the emitted $\gamma$ photon has
about a factor of 2 smaller $\gamma$ energy than a $\gamma$ photon in
electron beam direction. For angles $\Theta_{BW}= \frac{\sqrt{BW}}{\gamma_e}$,
the central $\gamma$ beam opening cone shows no longer the $\Theta$
dependence of the Compton scattering, but other factors like the band width
of the electron beam or the band width of the laser beam determine
the band width BW of the $\gamma$ beam. We will limit the acceptance
angle of the $\gamma$ lens or the corresponding $\gamma$ beam radius
$R$ at the entrance to the lens to this much smaller angle $\Theta_{BW}$.
Thus we make use of the very special $\gamma$ energy -- angle $\Theta$ 
correlation of Compton back-scattered $\gamma$ rays to work  with a very small
radius $R$ and to obtain a natural monochromatization. For a band width 
$BW=10^{-3}$, we thus expect for ELI-NP and MEGa-ray an intensity of $10^{10}\gamma$/s.
The outer angles $\Theta$ are no longer focused by the lens. 

\subsection{$\gamma$ Optics}

%\subsubsection{Refractive Optics}
\paragraph{Refractive Optics}

In the last 15 years, X-ray optics became very common at many synchrotron beam 
lines \cite{nielsen01,snigirev11}.
The refractive index of X-ray lenses is determined by the virtual 
photo effect (Rayleigh scattering) and follows the law \cite{lengeler99}:
\begin{equation}
  \label{eq:real}
 \delta_{photo}=-2.70 \cdot \frac{\lambda^2\cdot \rho \cdot Z}{A}\cdot 10^{-6}
       \propto \frac{1}{E_{\gamma}^2} \mbox{.}
\end{equation}
The $\gamma$ ray wave length $\lambda$ is measured in \AA, $\rho$ 
in $g/cm^3$, $Z$ is the atomic number and $A$ is the atomic mass in $g$. 
Obviously, $\delta_{photo}$ converges with $1/E_{\gamma}^2$ very fast 
towards zero, which represents a significant limitation for the 
construction of focusing optics at higher energies via $\delta_{photo}$. 
A way to still obtain a reasonable focal length $f$ is to combine a large 
number $N$ of lenses with 

\begin{equation}
  \label{eq:focal}
 f= R/(2\cdot\delta\cdot N),
 \end{equation}

where $R$ is the radius at the apex of the parabolically shaped, 
concave lenses.\\

However, recently \cite{habs11} we observed that the index of refraction
$\delta$ for $\gamma$ energies above 700 keV is positive in silicon (Z=14),
and no longer negative like for X-rays. For 1-2 MeV $\gamma$ rays we obtained
still rather small values of $\delta\approx +10^{-9}$. While for X-rays the 
virtual photo effect (Rayleigh scattering) explains $\delta$, for the higher
$\gamma$ energies it is the virtual pair creation or Delbr\"uck scattering.
Here no longer the electrons, but the electrostatic fields (beyond the Schwinger
 limit) of the nuclei dominate the scattering amplitude. 
Since the pair creation (Bethe-Heitler process) scales with $Z^2$,
we expect due the the Kramers-Kronig dispersion relation, that also
$\delta_{pair}$ contains in a power series development with $Z^2$
a term proportional to $Z^2$. But Delbr\"uck scattering, which scales
with $Z^4$, is for larger angles an absorptive process and
we expect from the Kramers-Kronig dispersion integral that 
$\delta_{pair}$ contains also a strong term scaling with $Z^4$. Extending the
same procedure to higher-order Delbr\"uck scattering and pair creation, 
$\delta_{pair}$ also contains in the expansion 
higher terms $Z^{2n}$. Even if the factors in front of $Z^{2n}$ are small,
for higher-Z materials these terms may start to dominate. Thus we expect
for typical high Z optical materials like gold or platinum, that the
index of refraction may reach values $\delta=10^{-5}-10^{-6}$ for 1-2 MeV
$\gamma$ rays. This large increase of $\delta$ with Z is essential
for all $\gamma$ optics and has to be verified experimentally. If larger
values of $\delta$ can be reached, smaller numbers of optical elements $N$
are required to obtain the same deflection. Also, the switch-over from 
negative to positive values of $\delta$ will occur for higher-Z materials
at lower $\gamma$ energy. For energies below, e.g., 100 keV, the large
increase of the photo effect with $Z^4$ probably makes high Z optical elements
inefficient. On the other hand, we have for gold above 0.5 MeV an absorption 
length $d_{1/2}$ of 3 mm, which is quite large.   

Thus, we expect for $\gamma$-beams that high-Z materials like gold or 
platinum will become the ideal material for all optical elements like 
lenses, prisms etc.. While for X rays due to the negative $\delta$
concave lenses are used for focusing, for $\gamma$ rays convex lenses
like in standard classical optics will be used.    
The new $\delta_{pair}$ shows a very different, much more favourable
$\gamma$ energy dependence: it is rather constant below 3 MeV and
drops off with  $1/E_{\gamma}$ for larger energies. Thus $\gamma$ optics 
may be extended up to 10-20 MeV. In contrast $\delta_{photo}$ showed the
much faster decrease with  $1/E_{\gamma}^2$ .
Typical $\gamma$ lenses at the beam entrance could have a 
radius of curvature of 500 $\mu$m, a central thickness of 150 $\mu$m 
and an acceptance diameter of 250 $\mu$m.

%\subsubsection{New Efficient Monochromator: Combination of
%               Refractive and Reflective $\gamma$ Optics}
\paragraph{New Efficient Monochromator: Combination of
                 Refractive and Reflective $\gamma$ Optics}

Monochromatization of electromagnetic radiation via Laue 
or Bragg diffraction follows Braggs law 

\begin{equation}
  \label{eq:mono}
      m \lambda = 2 d \sin \theta_B 
\end{equation}

Here $\lambda$ is the wave length of the incoming photon, 
$d$ the lattice spacing of the diffracting crystal,
$m$ is the order of reflection and $\theta_B$ the 
diffraction angle. We have a typical deflection 
angle of 10 mrad for 1 MeV $\gamma$ rays. This Bragg 
diffraction is a well established 
monochromatization technique \cite{nielsen01}.
A particular role in these considerations play 
so called perfect single crystals, where a 
coherent propagation of an incoming wave field all through the crystal occurs. 
Within dynamical diffraction theory \cite{nielsen01,rauch84} it can be shown
that the angular dependence of the intensity profile $I$, under which an 
incoming wave field is diffracted, has the following analytical 
form

\begin{eqnarray} 
I(\theta)&=&\sin^2\left(A\sqrt{1+y^2}\right)/(1+y^2) \label{e1} \\
y&\sim&(\theta-\theta_B)/{\lambda} \label{e2}
\end{eqnarray}

\begin{figure}
  \includegraphics[width=0.8 \textwidth]{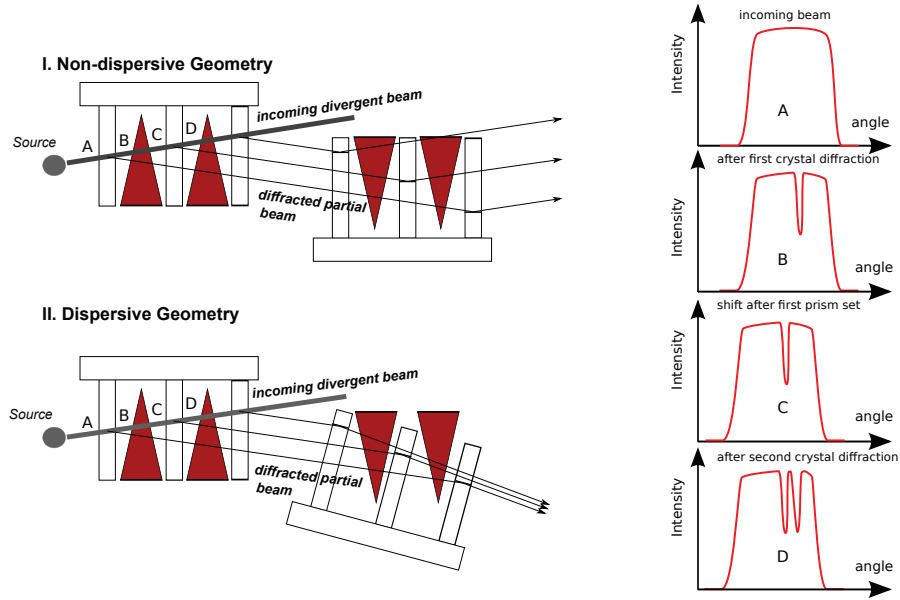}
  \caption{Schematic picture showing the successive small angular
           shifts between the sequence of double crystal monochromators.}
  \label{fig:mono1}
\end{figure}

Here $A$ is a crystal specific parameter 
which linearly depends on the wave length. It can be easily demonstrated 
from Eq.~(\ref{e1}) that the width of $I(\theta)$ is proportional 
to $\lambda$. This width should be called the acceptance 
width of the monochromator crystal. In order to obtain the excellent 
and wave-length independent possible monochromatization of an 
ideal crystal, it is necessary to create a beam with a divergence not 
larger than the acceptance width of the monochromator. This can be 
obtained by the use of a double-crystal monochromator 
\cite{kessler01,rauch84}. 
The beam accepted and diffracted by a first perfect crystal (in that sense 
the beam behind the first crystal) will have a divergence equal to the 
acceptance of this first crystal. If a second crystal, identical to the 
first one, is placed into this beam, the acceptance and divergence are 
perfectly matched and the theoretical possible monochromatization can 
be obtained. The second crystal operates at a different order of 
reflection $m$ for the dispersive mode, while the first crystal produces 
a narrow width in angle, which translates into a narrow energy width for 
the second crystal:

 \begin{equation}
  \label{eq:resolution}
 \Delta E_{\gamma}/E_{\gamma} 
= \Delta \Theta/\tan{\Theta}\approx  10 nrad/10 mrad\approx 10^{-6}
\end{equation}
This concept of monochromatization is largely used at X-ray synchrotrons,
 where the angular Darwin width is nicely matched to the natural opening 
angle $1/\gamma$. Since the acceptance of the first crystal  
quickly gets much smaller with wave length compared to the divergence 
of the incoming $\gamma$-beam, for $\gamma$-ray energies these intensity losses 
amount to 3-4 orders of magnitude. We want to pursue a 
new technical solution that allows to overcome these enormous 
losses in intensity associated with the use of perfect single 
crystals. The main idea consists in the combined use of refractive 
optics and diffraction monochromators and is 
illustrated in Fig. \ref{fig:mono1}. We propose to use instead of a single 
monochromator crystal a sequence of several crystals. To assure that 
each crystal has the same lattice spacing and the same orientation, 
we propose to cut them from a single large crystal block. In this sense 
each crystal sheet will act as a perfect monochromator crystal. 
Between the crystal sheets we propose to insert a system of one or several 
refracting prisms. The function of this combined system is illustrated 
in the right hand part of Fig. \ref{fig:mono1}. The upper graph (A) illustrates 
the divergent beam profile of the incoming photon beam. 
The next graph (B) shows the beam profile after the action of the 
first monochromator crystal sheet. It is indicated that a fraction 
of intensity is missing due to the diffraction process. In graph (C) we 
show how the prism system is acting: it is shifting the intensity profile 
in angular space (rotating it). By this it assures that the next 
monochromator crystal sees again the full intensity. The action of the 
next monochromator stage is shown in plot (D). It is important to notice 
that the action of the refractive prism should shift the intensity 
as much as the acceptance width of each single crystal sheet would be.
A sequence of crystal sheets and refractive prisms is supposed to replace 
a single monochromator crystal within a double-crystal monochromator setup. 
The parallel single crystal slabs require a certain thickness, depending
on the photon energy but typically 1-5 mm. Due to the negligible 
absorption in the crystals and wedges, the process can be repeated typically 
10 -100 times. Since we will first match the incoming beam with a $\gamma$ lens
system to the monochromator system, we expect in the range of 0.1 - 20 MeV 
a further improvement in the order of 10 -100, getting
close to a reduction in intensity just by the factor of $10^3$ corresponding
to the ratio of initial energy resolution $(10^{-3})$ to the final energy
resolution $(10^{-6})$.

\section{Applications}
\begin{figure}
  \includegraphics[width=0.6 \textwidth]{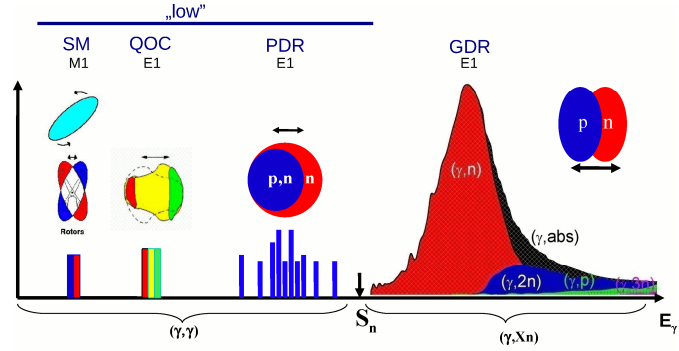}
  \caption{Nuclear M1 and E1 excitations. SM= Scissors Mode, QOC=
           Quadrupole-Octupole Mode, PDR= Pygmy Dipole Resonance, 
           GDR= Giant Dipole Resonance.}
  \label{fig:nuclear}
\end{figure}

In the discussion of possible $\gamma$-beam applications 
we first consider the cross section of photonuclear reactions
and the width $\Gamma_{\gamma}$. The cross section for a compound
nucleus resonance populated by photo-excitation
at the resonance energy $E_r$ in the region below
the neutron separation energy $S_n$ is given by the Breit-Wigner
formula \cite{segre77}

\begin{equation}
   \sigma (E_{\gamma})= (\lambda_{\gamma}^2/4\pi)\cdot g \cdot
   \frac{\Gamma_{\gamma}\Gamma_2}{(E_{\gamma}-E_r)^2 +
   (\Gamma)^2/4},
\end{equation}

where $g=\frac{2I_a+1}{2I_b+1}$ is a spin factor for the spin of
the target and the beam and $\lambda_{\gamma}=(h \cdot
c)/E_{\gamma}$ represents the wavelength of the $\gamma$ rays with
energy $E_{\gamma}$. The resonance is excited with the width
$\Gamma_{\gamma}$ and decays via the width $\Gamma_2$. The
resonance has a total width $\Gamma=\Gamma_{\gamma}+\Gamma_2$.
Fig. \ref{fig:nuclear} shows the dominant E1 and M1 nuclear excitation 
widths $\Gamma_{\gamma}$.

The typical maximum E1 width for a level up to the neutron separation 
energy $S_n$ is about 10 milli-Weisskopf units. It corresponds for 2 MeV
photons to about 0.1 eV. On the other hand,   
the Doppler broadening of a $\gamma$ transition at room temperature
$kT=1/40$ eV for a nucleus with mass number $A=200$ and a $\gamma$ energy
$E_{\gamma}$= 2 MeV is

\begin{equation}
   \Delta E_{\gamma}=E_{\gamma}\sqrt{(2kT)/m_pc^2A}\approx \mbox{1~eV}.
\end{equation}

It shows that we find a typical relative line width of $10^{-6}$ for a 
nucleus with a mass number A=50 or 1 eV at 1 MeV and 10 eV at 10 MeV, 
respectively. Thus with a $\gamma$
beam of $10^{-6}$ band width we can selectively address individual nuclear 
levels within their Doppler width. Since the maximum nuclear cross sections
$\lambda_{\gamma}^2/4\pi$ are typically 1 kbarn for 1 MeV and thus much larger
than atomic cross sections of 1 -10 barn for Compton scattering and pair
creation, we can strongly reduce the atomic background. At the same time 
the deexcitation cascades of individual levels are much less complex than
when exciting a multitude of levels at the same time. 

\begin{figure}
  \includegraphics[width=0.9 \textwidth]{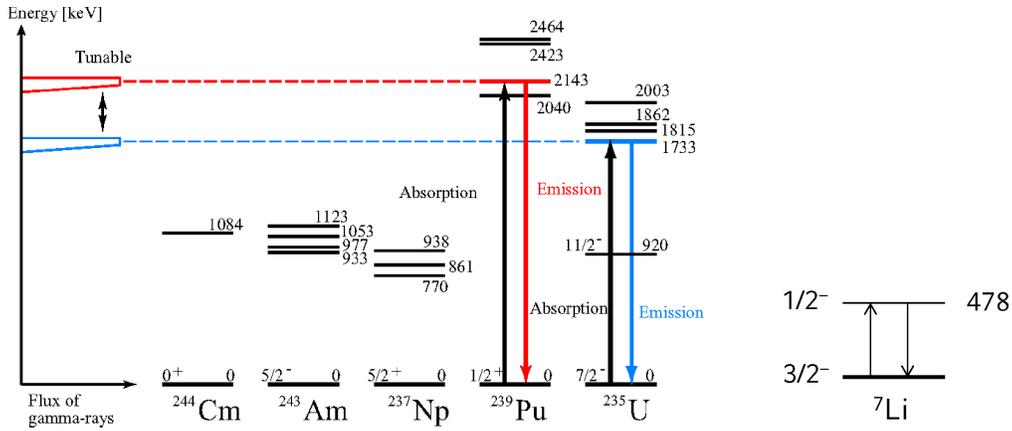}
  \caption{Nuclear resonance fluorescence M1 excitation modes in actinides.}
  \label{fig:M1}
\end{figure}
 
Now, let us consider the M1 and E1 excitations of Fig. \ref{fig:nuclear} as
strong excitations for potential applications. The low-lying M1 excitation 
can be employed via Nuclear Resonance Fluorescence (NRF) using focusing 
lenses for tomography and microscopy with $\mu$m resolution \cite{schroer10}. 
In Fig. \ref{fig:M1} 
we show M1 excitations for actinides, 
which can be used in nuclear fuel assay or radioactive waste monitoring.
Based on the M1 transition in $^7$Li (478 keV) we can, e.g., perform a 
high resolution 
tomography of growing defects in Li batteries or in medical physics we can study
 the distribution of Li in the brain during treatment of manic depressive
alternating insanity.

Due to their internal structure, the higher QOC E1 excitations frequently decay
by quadrupole transitions to higher-spin states. We want to use this feature
to populate high-spin isomers like $^{195m}$Pt to produce medical radioisotopes for
diagnostics and therapy \cite{koester11}.

We envisage to use the pygmy resonance to populate very weakly bound neutron states
just below $S_n$ from the stable ground state by an E1 excitation. These
weakly bound states are isomers, because the neutron tunnels out very far.
With a second laser we can release this neutron, generating a very brilliant
neutron source \cite{boenia11}. Also a very brilliant secondary positron 
source can be produced from high-Z foils of e.g. platinum with
a 1.5-2.0 MeV $\gamma$-beam via the Bethe-Heitler pair creation process
\cite{hugenschmidt11}. We also proposed the production of many new
medical radioisotopes via the giant dipole resonance \cite{koester11}.
Compared to X rays, one of the main advantages of $\gamma$ rays is their
deep penetration. With the new $\gamma$ optics many new isotope specific
tomographic studies become possible.

In the future we plan to increase the intensity of very monochromatic,
very directed Compton back-scattered $\gamma$-beams by a factor of
about $10^6$, seeding the emission of a $\gamma$ quantum
FEL \cite{bonifacio11,bonifacio09,guenther11}. This would result in the
possibility of many large scale industrial applications, but also
a new field of coherent $\gamma$-ray physics \cite{paganin06} with 
much larger coupling strength would open up.

%%%%%%%%%%%%%%%%%%%%%%%%%%%%%%%%%%%%%%%%%%%%%%%%
%% BACKMATTER
%%%%%%%%%%%%%%%%%%%%%%%%%%%%%%%%%%%%%%%%%%%%%%%%

%\begin{theacknowledgments}
  
%\end{theacknowledgments}

%%%%%%%%%%%%%%%%%%%%%%%%%%%%%%%%%%%%%%%%%%%%%%%%
%% The bibliography can be prepared using the BibTeX program or
%% manually.
%%
%% The code below assumes that BibTeX is used.  If the bibliography is
%% produced without BibTeX comment out the following lines and see the
%% aipguide.pdf for further information.
%%
%% For your convenience a manually coded example is appended
%% after the \end{document}
%%%%%%%%%%%%%%%%%%%%%%%%%%%%%%%%%%%%%%%%%%%%%%%%

%%%%%%%%%%%%%%%%%%%%%%%%%%%%%%%%%%%%%%%%%%%%%%%%
%% You may have to change the BibTeX style below, depending on your
%% setup or preferences.
%%
%%
%% For The AIP proceedings layouts use either
%%%%%%%%%%%%%%%%%%%%%%%%%%%%%%%%%%%%%%%%%%%%

\bibliographystyle{aipproc}   % if natbib is available
%\bibliographystyle{aipprocl} % if natbib is missing

%%%%%%%%%%%%%%%%%%%%%%%%%%%%%%%%%%%%%%%%%%%
%% You probably want to use your own bibtex database here
%%%%%%%%%%%%%%%%%%%%%%%%%%%%%%%%%%%%%%%%%%%
%\bibliography{sample}

%%%%%%%%%%%%%%%%%%%%%%%%%%%%%%%%%%%%%%%%%%%
%% Just a reminder that you may have to run bibtex
%% All of it up to \end{document} can be removed
%% if you don't like the warning.
%%%%%%%%%%%%%%%%%%%%%%%%%%%%%%%%%%%%%%%%%%%
%\IfFileExists{\jobname.bbl}{}
 %{\typeout{}
  %\typeout{******************************************}
  %\typeout{** Please run "bibtex \jobname" to optain}
  %\typeout{** the bibliography and then re-run LaTeX}
  %\typeout{** twice to fix the references!}
  %\typeout{******************************************}
  %\typeout{}
 %}

%%%%%%%%%%%%%%%%%%%%%%%%%%%%%%%%%%%%%%%%%%%
%% The following lines show an example how to produce a bibliography
%% without the help of the BibTeX program. This could be used instead
%% of the above.
%%%%%%%%%%%%%%%%%%%%%%%%%%%%%%%%%%%%%%%%%%%

%\endinput

\end{document}